%Paper: hep-ph/9412385
%From: ramond@phys.ufl.edu
%Date: Thu, 29 Dec 94 12:21:21 -0500

%      Page layout, margins (feel free to change)
\hsize=6.5truein
\hoffset=.3truein
\vsize=8.9truein
\voffset=.3truein
%%%%%%%%%%%%%%%%%%%%%%%%%%%%%%%%%%%%%%%%%%%%%%%%%%%%%%%%%%%%%%%%%%%%%%%%%%%%%
% This is sjnl.tex, by Doug Eardley's jnl, eqnorder and reforder, %
% but combined, modified and reduced to order      %
%%%%%%%%%%%%%%%%%%%%%%%%%%%%%%%%%%%%%%%%%%%%%%%%%%%%%%%%%%%%%%%%%%%%%%%%%%%%%
%  Define pseudo-12pt fonts
\font\twelverm=cmr10 scaled 1200    \font\twelvei=cmmi10 scaled 1200
\font\twelvesy=cmsy10 scaled 1200   \font\twelveex=cmex10 scaled 1200
\font\twelvebf=cmbx10 scaled 1200   \font\twelvesl=cmsl10 scaled 1200
\font\twelvett=cmtt10 scaled 1200   \font\twelveit=cmti10 scaled 1200
\skewchar\twelvei='177   \skewchar\twelvesy='60
%  Define \...point macros to change fonts and spacings consistently
%\def\twelvepoint{\normalbaselineskip=12.4pt
\def\twelvepoint{\normalbaselineskip=14pt
  \abovedisplayskip 12.4pt plus 3pt minus 9pt
  \belowdisplayskip 12.4pt plus 3pt minus 9pt
  \abovedisplayshortskip 0pt plus 3pt
  \belowdisplayshortskip 7.2pt plus 3pt minus 4pt
  \smallskipamount=3.6pt plus1.2pt minus1.2pt
  \medskipamount=7.2pt plus2.4pt minus2.4pt
  \bigskipamount=14.4pt plus4.8pt minus4.8pt
  \def\rm{\fam0\twelverm}          \def\it{\fam\itfam\twelveit}%
  \def\sl{\fam\slfam\twelvesl}     \def\bf{\fam\bffam\twelvebf}%
  \def\mit{\fam 1}                 \def\cal{\fam 2}%
  \def\tt{\twelvett}
  \textfont0=\twelverm   \scriptfont0=\tenrm   \scriptscriptfont0=\sevenrm
  \textfont1=\twelvei    \scriptfont1=\teni    \scriptscriptfont1=\seveni
  \textfont2=\twelvesy   \scriptfont2=\tensy   \scriptscriptfont2=\sevensy
  \textfont3=\twelveex   \scriptfont3=\twelveex  \scriptscriptfont3=\twelveex
  \textfont\itfam=\twelveit
  \textfont\slfam=\twelvesl
  \textfont\bffam=\twelvebf \scriptfont\bffam=\tenbf
  \scriptscriptfont\bffam=\sevenbf
  \normalbaselines\rm}
%       tenpoint

%      Various internal macros
\def\beginlinemode{\endmode
  \begingroup\parskip=0pt \obeylines\def\\{\par}\def\endmode{\par\endgroup}}
\def\beginparmode{\endmode
  \begingroup \def\endmode{\par\endgroup}}
\let\endmode=\par
{\obeylines\gdef\
{}}
\def\singlespace{\baselineskip=\normalbaselineskip}
\def\oneandahalfspace{\baselineskip=\normalbaselineskip
  \multiply\baselineskip by 3 \divide\baselineskip by 2}
\def\doublespace{\baselineskip=\normalbaselineskip \multiply\baselineskip by 2}
\newcount\firstpageno
\firstpageno=2
\footline={\ifnum\pageno<\firstpageno{\hfil}\else{\hfil\twelverm\folio\hfil}\fi}
\let\rawfootnote=\footnote              % We must set the footnote style
\def\footnote#1#2{{\rm\singlespace\parindent=0pt\rawfootnote{#1}{#2}}}
\def\raggedcenter{\leftskip=2em plus 12em \rightskip=\leftskip
  \parindent=0pt \parfillskip=0pt \spaceskip=.3333em \xspaceskip=.5em
  \pretolerance=9999 \tolerance=9999
  \hyphenpenalty=9999 \exhyphenpenalty=9999 }
\parskip=\medskipamount
\twelvepoint            % selects twelvepoint fonts (cf. \tenpoint)
\overfullrule=0pt       % delete the nasty little black boxes for overfull box
\def\preprintno#1{
 \rightline{\rm #1}}    % Preprint number at upper right of title page
\def\author                     %  Author(s) name(s)  on title page
  {\vskip 3pt plus 0.2fill \beginlinemode
   \singlespace \raggedcenter \twelvesc}
\def\affil                      % Affiliations (can intermix with \author)
  {\vskip 3pt plus 0.1fill \beginlinemode
   \oneandahalfspace \raggedcenter \sl}
\def\abstract                   % Begin abstract
  {\vskip 3pt plus 0.3fill \beginparmode
   \doublespace \narrower \noindent ABSTRACT: }
\def\endtitlepage               % End title page, begin body of paper
  {\endpage                     %       This subsumes \body
   \body}
\def\body                       % Begin text body;  can be used to end
  {\beginparmode}               % \title, \author, \affil, \abstract,
                                % \reference, or \figurecaption modes

%%\def\subhead#1{                 % Subhead;  NOTE enclose the text in {}
%%  \vskip 0.25truein             % e.g., \subhead{A. History of the Problem}
%%  {\raggedcenter #1 \par}
%%   \nobreak\vskip 0.25truein\nobreak}
\def\subhead#1{                 % Subhead;  NOTE enclose the text in {}
  \vskip 0.1truein             % e.g., \subhead{A. History of the Problem}
  {\raggedcenter #1 \par}
   \nobreak\vskip 0.1truein\nobreak}
\def\refto#1{$|{#1}$}           % For references in text as superscript
\def\references                 % Begin references -- basic format is Phys Rev
  {\subhead{References}         % I.e., volume, page, year (space after
%%commas).
   \beginparmode
   \frenchspacing \parindent=0pt \leftskip=1truecm
   \parskip=8pt plus 3pt \everypar{\hangindent=\parindent}}
\gdef\refis#1{\indent\hbox to 0pt{\hss#1.~}}    % Ref list numbers.
\gdef\journal#1, #2, #3, 1#4#5#6{               % Journal reference.  Comma
%%sets
    {\sl #1~}{\bf #2}, #3, (1#4#5#6)}           % off: name, vol, page, year
\def\refstylenp{                % Nucl Phys(or Phys Lett) ref style: V, Y, P
  \gdef\refto##1{ [##1]}                                % Reference in text []
  \gdef\refis##1{\indent\hbox to 0pt{\hss##1)~}}        % Ref list numbers)
  \gdef\journal##1, ##2, ##3, ##4 {                     % Journal reference
     {\sl ##1~}{\bf ##2~}(##3) ##4 }}
\def\refstyleprnp{              % Input like pr, output like np!!
  \gdef\refto##1{ [##1]}                                % Reference in text []
  \gdef\refis##1{\indent\hbox to 0pt{\hss##1)~}}        % Ref list numbers)
  \gdef\journal##1, ##2, ##3, 1##4##5##6{               % Journal reference
    {\sl ##1~}{\bf ##2~}(1##4##5##6) ##3}}
\def\pr{\journal Phys. Rev., }

\def\prl{\journal Phys. Rev. Lett., }
\def\prpts{\journal Phys. Rep., }
\def\np{\journal Nucl. Phys., }
\def\pl{\journal Phys. Lett., }

\def\endreferences{\body}
\def\endpage                    %  Eject a page
  {\vfill\eject}
\def\endpaper                   %  Ways to say goodbye
  {\endmode\vfill\supereject}
\def\endit
  {\endpaper\end}
%%      Various user definitions
\def\ref#1{Ref. #1}                     %       for inline references
\def\Ref#1{Ref. #1}                     %       ditto

\def\m@th{\mathsurround=0pt }
\font\twelvesc=cmcsc10 scaled 1200
\def\cite#1{{#1}}
\def\(#1){(\call{#1})}
\def\call#1{{#1}}
\def\taghead#1{}
\def\leaderfill{\leaders\hbox to 1em{\hss.\hss}\hfill}
\def\twiddle{\lower.9ex\rlap{$\kern-.1em\scriptstyle\sim$}}
\def\bigtwiddle{\lower1.ex\rlap{$\sim$}}
\def\gtwid{\mathrel{\raise.3ex\hbox{$>$\kern-.75em\lower1ex\hbox{$\sim$}}}}
\def\ltwid{\mathrel{\raise.3ex\hbox{$<$\kern-.75em\lower1ex\hbox{$\sim$}}}}
\def\square{\kern1pt\vbox{\hrule height 1.2pt\hbox{\vrule width 1.2pt\hskip 3pt
   \vbox{\vskip 6pt}\hskip 3pt\vrule width 0.6pt}\hrule height 0.6pt}\kern1pt}
%%		EQNORDER.TEX			11/05/85	Doug E.
\catcode`@=11
\newcount\tagnumber\tagnumber=0

\immediate\newwrite\eqnfile
\newif\if@qnfile\@qnfilefalse
\def\write@qn#1{}
\def\writenew@qn#1{}
\def\w@rnwrite#1{\write@qn{#1}\message{#1}}
\def\@rrwrite#1{\write@qn{#1}\errmessage{#1}}

\def\taghead#1{\gdef\t@ghead{#1}\global\tagnumber=0}
\def\t@ghead{}

\expandafter\def\csname @qnnum-3\endcsname
  {{\t@ghead\advance\tagnumber by -3\relax\number\tagnumber}}
\expandafter\def\csname @qnnum-2\endcsname
  {{\t@ghead\advance\tagnumber by -2\relax\number\tagnumber}}
\expandafter\def\csname @qnnum-1\endcsname
  {{\t@ghead\advance\tagnumber by -1\relax\number\tagnumber}}
\expandafter\def\csname @qnnum0\endcsname
  {\t@ghead\number\tagnumber}
\expandafter\def\csname @qnnum+1\endcsname
  {{\t@ghead\advance\tagnumber by 1\relax\number\tagnumber}}
\expandafter\def\csname @qnnum+2\endcsname
  {{\t@ghead\advance\tagnumber by 2\relax\number\tagnumber}}
\expandafter\def\csname @qnnum+3\endcsname
  {{\t@ghead\advance\tagnumber by 3\relax\number\tagnumber}}

\def\equationfile{%
  \@qnfiletrue\immediate\openout\eqnfile=\jobname.eqn%
  \def\write@qn##1{\if@qnfile\immediate\write\eqnfile{##1}\fi}
  \def\writenew@qn##1{\if@qnfile\immediate\write\eqnfile
    {\noexpand\tag{##1} = (\t@ghead\number\tagnumber)}\fi}
}

\def\callall#1{\xdef#1##1{#1{\noexpand\call{##1}}}}
\def\call#1{\each@rg\callr@nge{#1}}

\def\each@rg#1#2{{\let\thecsname=#1\expandafter\first@rg#2,\end,}}
\def\first@rg#1,{\thecsname{#1}\apply@rg}
\def\apply@rg#1,{\ifx\end#1\let\next=\relax%
\else,\thecsname{#1}\let\next=\apply@rg\fi\next}

\def\callr@nge#1{\calldor@nge#1-\end-}
\def\callr@ngeat#1\end-{#1}
\def\calldor@nge#1-#2-{\ifx\end#2\@qneatspace#1 %
  \else\calll@@p{#1}{#2}\callr@ngeat\fi}
\def\calll@@p#1#2{\ifnum#1>#2{\@rrwrite{Equation range #1-#2\space is bad.}
\errhelp{If you call a series of equations by the notation M-N, then M and
N must be integers, and N must be greater than or equal to M.}}\else%
 {\count0=#1\count1=#2\advance\count1
by1\relax\expandafter\@qncall\the\count0,%
  \loop\advance\count0 by1\relax%
    \ifnum\count0<\count1,\expandafter\@qncall\the\count0,%
  \repeat}\fi}

\def\@qneatspace#1#2 {\@qncall#1#2,}
\def\@qncall#1,{\ifunc@lled{#1}{\def\next{#1}\ifx\next\empty\else
  \w@rnwrite{Equation number \noexpand\(>>#1<<) has not been defined yet.}
  >>#1<<\fi}\else\csname @qnnum#1\endcsname\fi}

\let\eqnono=\eqno
\def\eqno(#1){\tag#1}
\def\tag#1$${\eqnono(\displayt@g#1 )$$}

\def\aligntag#1\endaligntag
  $${\gdef\tag##1\\{&(##1 )\cr}\eqalignno{#1\\}$$
  \gdef\tag##1$${\eqnono(\displayt@g##1 )$$}}

\def\eqalignno#1{\displ@y \tabskip\centering
  \halign to\displaywidth{\hfil$\displaystyle{##}$\tabskip\z@skip
    &$\displaystyle{{}##}$\hfil\tabskip\centering
    &\llap{$\displayt@gpar##$}\tabskip\z@skip\crcr
    #1\crcr}}

\def\displayt@gpar(#1){(\displayt@g#1 )}

\def\displayt@g#1 {\rm\ifunc@lled{#1}\global\advance\tagnumber by1
        {\def\next{#1}\ifx\next\empty\else\expandafter
        \xdef\csname @qnnum#1\endcsname{\t@ghead\number\tagnumber}\fi}%
  \writenew@qn{#1}\t@ghead\number\tagnumber\else
        {\edef\next{\t@ghead\number\tagnumber}%
        \expandafter\ifx\csname @qnnum#1\endcsname\next\else
        \w@rnwrite{Equation \noexpand\tag{#1} is a duplicate number.}\fi}%
  \csname @qnnum#1\endcsname\fi}

\def\ifunc@lled#1{\expandafter\ifx\csname @qnnum#1\endcsname\relax}

\let\@qnend=\end\gdef\end{\if@qnfile
\immediate\write16{Equation numbers written on []\jobname.EQN.}\fi\@qnend}

\catcode`@=12
%%%%%%%%%%%%%%%%%%%%             REFORDER.TEX              %%%%%%%%%%%%%%%%%%%%
\refstyleprnp
\catcode`@=11
\newcount\r@fcount \r@fcount=0
\def\refreset{\global\r@fcount=0}
\newcount\r@fcurr
\immediate\newwrite\reffile
\newif\ifr@ffile\r@ffilefalse
\def\w@rnwrite#1{\ifr@ffile\immediate\write\reffile{#1}\fi\message{#1}}

\def\writer@f#1>>{}
\def\referencefile{%			  Stuff to write .REF file
  \r@ffiletrue\immediate\openout\reffile=\jobname.ref%
  \def\writer@f##1>>{\ifr@ffile\immediate\write\reffile%
    {\noexpand\refis{##1} = \csname r@fnum##1\endcsname = %
     \expandafter\expandafter\expandafter\strip@t\expandafter%
     \meaning\csname r@ftext\csname r@fnum##1\endcsname\endcsname}\fi}%
  \def\strip@t##1>>{}}

\def\citeall#1{\xdef#1##1{#1{\noexpand\cite{##1}}}}
\def\cite#1{\each@rg\citer@nge{#1}}	% Variable No. of args, separated by ","

\def\each@rg#1#2{{\let\thecsname=#1\expandafter\first@rg#2,\end,}}
\def\first@rg#1,{\thecsname{#1}\apply@rg}	% each@ag is a general purpose
\def\apply@rg#1,{\ifx\end#1\let\next=\relax%	  variable no. of arg. macro.
\else,\thecsname{#1}\let\next=\apply@rg\fi\next}% args separated by commas

\def\citer@nge#1{\citedor@nge#1-\end-}	% Check for M-N range (M and N numbers)
\def\citer@ngeat#1\end-{#1}
\def\citedor@nge#1-#2-{\ifx\end#2\r@featspace#1 % Single argument
  \else\citel@@p{#1}{#2}\citer@ngeat\fi}	% M-N range of arguments
\def\citel@@p#1#2{\ifnum#1>#2{\errmessage{Reference range #1-#2\space is bad.}%
    \errhelp{If you cite a series of references by the notation M-N, then M and
    N must be integers, and N must be greater than or equal to M.}}\else%
 {\count0=#1\count1=#2\advance\count1
by1\relax\expandafter\r@fcite\the\count0,%
  \loop\advance\count0 by1\relax%	  Loop from M to N
    \ifnum\count0<\count1,\expandafter\r@fcite\the\count0,%
  \repeat}\fi}

\def\r@featspace#1#2 {\r@fcite#1#2,}	% Eat spaces at beginning or end of arg
\def\r@fcite#1,{\ifuncit@d{#1}%		  Cite individual reference
    \newr@f{#1}%
    \expandafter\gdef\csname r@ftext\number\r@fcount\endcsname%
                     {\message{Reference #1 to be supplied.}%
                      \writer@f#1>>#1 to be supplied.\par}%
 \fi%
 \csname r@fnum#1\endcsname}
\def\ifuncit@d#1{\expandafter\ifx\csname r@fnum#1\endcsname\relax}%
\def\newr@f#1{\global\advance\r@fcount by1%
    \expandafter\xdef\csname r@fnum#1\endcsname{\number\r@fcount}}

\let\r@fis=\refis			% Save old \refis, redefine
\def\refis#1#2#3\par{\ifuncit@d{#1}%      Use two params #2 #3 to strip blank
   \newr@f{#1}%
   \w@rnwrite{Reference #1=\number\r@fcount\space is not cited up to now.}\fi%
  \expandafter\gdef\csname r@ftext\csname r@fnum#1\endcsname\endcsname%
  {\writer@f#1>>#2#3\par}}

\def\ignoreuncited{%   redefine \refis if ignoring uncited references
   \def\refis##1##2##3\par{\ifuncit@d{##1}%
     \else\expandafter\gdef\csname r@ftext\csname
r@fnum##1\endcsname\endcsname%
     {\writer@f##1>>##2##3\par}\fi}}

\def\r@ferr{\endreferences\errmessage{I was expecting to see
\noexpand\endreferences before now;  I have inserted it here.}}
\let\r@ferences=\references
\def\references{\r@ferences\def\endmode{\r@ferr\par\endgroup}}

\let\endr@ferences=\endreferences
\def\endreferences{\r@fcurr=0%		  Save old \endreferences, redefine
  {\loop\ifnum\r@fcurr<\r@fcount%	  Loop over refnum and produce text
    \advance\r@fcurr by 1\relax\expandafter\r@fis\expandafter{\number\r@fcurr}%
    \csname r@ftext\number\r@fcurr\endcsname%
  \repeat}\gdef\r@ferr{}\global\r@fcount=0\endr@ferences}

\let\r@fend=\endpaper\gdef\endpaper{\ifr@ffile
\immediate\write16{Cross References written on []\jobname.REF.}\fi\r@fend}

\catcode`@=12

\citeall\refto		% These macros will generate citations
\citeall\ref		%
\citeall\Ref		%

\def\oneandthreefifthsspace{\baselineskip=\normalbaselineskip
  \multiply\baselineskip by 8 \divide\baselineskip by 5}

\font\titlefont=cmr10 scaled\magstep3
\def\bigtitle                      %  Title on title page
  {\null\vskip 3pt plus 0.2fill
   \beginlinemode \doublespace \raggedcenter \titlefont}

%%%%%%%%%%%%%%%%%%%%%%%%%%%%%%%%%%%%%%%%%%%%%%%%%%%%%%%%%%%%%%%%%%%%%%%%%%%%%%%
\def\uof{Institute for Fundamental Theory\\Department of Physics,
University of Florida\\Gainesville FL 32611}
\def\orsay{Laboratoire de Physique Th\' eorique et Hautes
Energies\footnote*{Laboratoire Associ\'e au Centre National de la
Recherche Scientifique\hfill}
\\B\^atiment 211\\F--91405 Orsay CEDEX,  France}
%%%%%%%%%%%%%%%%%%%%%%%%%%%%%%%%%%%%%%%%%%%%%%%%%%%%%%%%%%%%%%%%%%%%%%%%%%%%%%%
\preprintno{LPTHE-ORSAY 94/115}
\preprintno{UFIFT-HEP-94-19}
\preprintno{final 12-29-94}
\bigtitle{Yukawa Textures and Anomalies}
\author Pierre Bin\'etruy
\affil\orsay
\vskip.3cm
\centerline{and}
\author Pierre Ramond\footnote{**}{Supported in part by the United States
Department of Energy under grant DE-FG05-86-ER40272}
\affil\uof
\body
\abstract
We augment the Minimal Supersymmetric Standard Model with a gauged
family-dependent $U(1)$ to reproduce Yukawa textures compatible with
experiment. In the simplest model with one extra chiral electroweak
singlet field, acceptable  textures require this  $U(1)$ to be
anomalous. The cancellation of its anomalies  by a generic Green-Schwarz
mechanism requires $\sin^2\theta_w=3/8$ at the string scale, suggesting a
superstring origin for the standard model.
\endtitlepage
\oneandthreefifthsspace

\subhead{1.~\bf Introduction}
\taghead{1.}

The extension of the standard model to $N=1$
supersymmetry[\cite{reviews}] allows for its perturbative extrapolation
to near Planckian scales, where the gauge couplings
[\cite{unification}]and some Yukawa couplings[\cite{btau}] appear to
converge. This raises the hope that the $N=1$ standard model at short
distances is much simpler than at experimental scales. However we do not
have sufficient information to determine exactly the type of structure
it describes, a GUT theory[\cite{gut}], or a direct descendant of
superstrings.

In this letter, we attempt to answer this question by considering the
structure of the Yukawa couplings. While not known in detail, their
orders of magnitude are well determined by experiment. Their most
striking aspect is the hierarchy of the masses of the three chiral
families. The experimental values of the quark and lepton masses,
extrapolated near the Planck scale, satisfy  the orders of magnitude
estimates[\cite{RRR}]
$${m_u\over m_t}={\cal O}(\lambda^8)\ ;\qquad {m_c\over m_t}={\cal
O}(\lambda^4)\ ; \eqno(top)$$
$${m_d\over m_b}={\cal
O}(\lambda^4)\ ; \qquad {m_s\over m_b}={\cal
O}(\lambda^2)\ ,\eqno(bottom)$$
where, following Wolfenstein's parametrization[\cite{wolf}], we use the
Cabibbo angle $\lambda$, as expansion parameter. The charged lepton
masses also satisfy similar relations
$${m_e\over m_\tau}={\cal
O}(\lambda^4)\ ; \qquad {m_\mu\over m_\tau}={\cal
O}(\lambda^2)\ .\eqno(lept)$$
The mass hierarchy appears to be geometrical in each sector. The equality
$$m_b=m_\tau\ ,$$
known to be valid in the ultraviolet[\cite{btau}], yields the estimate
$${m_dm_sm_b\over m_e m_\mu m_\tau}={\cal O}(1) \ .\eqno(goodrat)$$

A question of intense theoretical speculation is the mechanism which
sets these orders of magnitude. In this letter we  explore the
possibility that it is a family-dependent gauged Abelian
symmetry[\cite{WORKSHOP}]. While
hardly new, this idea has been revisited in the recent literature
[\cite{LNS, IR,PAPAG}], but in rather specific models. We find in the
simplest model of this kind that in order to reproduce the orders of
magnitude of the quarks and lepton masses, the Abelian family symmetry
must be anomalous. When its anomaly is compensated by a Green-Schwarz
mechanism, the Weinberg angle is fixed[\cite{Ib}]. Remarkably its value
is $\sin^2\theta_w=3/8$, in perfect agreement with data when
extrapolated to the infrared.

Our framework is the minimal extension of the Standard Model to $N=1$
supersymmetry, including the so-called $\mu$ term, $P=\mu H_uH_d\ .$ We
reserve to a forthcoming paper[\cite{BLR}] the case without an {\it ab
initio} $\mu$ term. In this paper, we  aim to determine to what extent
an Abelian charge symmetry can help in narrowing down possible Yukawa
textures.

\subhead{2.~\bf Yukawa Textures}
\taghead{2.}
Consider the most general Abelian charge that can be assigned to
the particles of the Supersymmetric Standard Model,
$$X=X^{}_0+X^{}_3+{\sqrt 3}X^{}_8\ ,\eqno(X)$$
where $X_0$ is the family-independent part, $X_3$ is along $\lambda_3$,
and $X_8$ is along $\lambda_8$, the two diagonal Gell-Mann matrices of
the $SU(3)$ family space in each charge sector. In a basis where the
entries correspond to the components in the family space of the fields
${\bf Q}$, $\overline{\bf u}$, $\overline{\bf d}$, $L$, and $\overline
e$, we can write the different components in the form
$$X_i^{}=(a^{}_i,b^{}_i,c_i^{},d_i^{},e_i^{})\ ,\eqno(Xfamily)$$
for $i=0,3,8$.
The Higgs doublets $H_{u,d}$ have zero X-charge because of the $\mu$
term.

Let us assume that the tree-level Yukawa coupling involves {\it only}
the third family (implicitly choosing the third direction in family
space),
$$y_t{\bf Q}_3\overline{\bf u}_3H_u+y_b{\bf Q}_3\overline{\bf d}_3H_d+
y_\tau L_3\overline e_3H_d\ ,\eqno(Yuk)$$
where the $y_i$'s are the Yukawa couplings. This generates the relations
$$\eqalign{{a_0^{}+b_0^{}}=&2(a_8+b_8)\ ,\cr
{a_0^{}+c_0^{}}=&2(a_8+c_8)\ ,\cr
{d_0^{}+e_0^{}}=&2(d_8+e_8)\ .\cr}\eqno(Yuky)$$
The other elements of the Yukawa matrices are zero at tree-level. We
assume that the reason is conservation of X-charge: these entries do not
have the correct X-charge for renormalizable couplings.  Let $x_{ij}$
be the  excess X-charges at each of their entries; they are

$${\rm charge}~{2\over 3}:
\pmatrix{3(a_8+b_8)+a_3+b_3&3(a_8+b_8)+a_3-b_3&3a_8+a_3\cr
3(a_8+b_8)-a_3+b_3&
3(a_8+b_8)-a_3-b_3&3a_8-a_3\cr
3b_8+b_3&3b_8-b_3&0\cr}$$

$${\rm charge}~-{1\over 3}:
\pmatrix{3(a_8+c_8)+a_3+c_3&3(a_8+c_8)+a_3-c_3&3a_8+a_3\cr
3(a_8+c_8)-a_3+c_3&
3(a_8+c_8)-a_3-c_3&3a_8-a_3\cr
3c_8+c_3&3c_8-c_3&0\cr}$$

$${\rm charge}~-1:
\pmatrix{3(d_8+e_8)+d_3+e_3&3(d_8+e_8)+d_3-e_3&3d_8+d_3\cr
3(d_8+e_8)-d_3+e_3&
3(d_8+e_8)-d_3-e_3&3d_8-d_3\cr
3e_8+e_3&3e_8-e_3&0\cr}$$
For the empty entries in the Yukawa matrices to be filled, the
tree-level chiral symmetry of the first two families must be broken. A
generic mechanism is to couple the fields of the standard model with new
vector-like fermions, and then give these new fermions a mass that
breaks chiral symmetry, and sets the scale of chiral symmetry breaking.
This is similar to the see-saw mechanism[\cite{seesaw}], where chiral
symmetry is replaced by lepton number, but its implementation is different,
since we deal with Dirac, rather than Majorana matrices. It is a matter
for the model builder to offer specific models which realize this
scenario; we assume it can be done, and refer the reader to models in
the literature[\cite{FN,hall}].

The excess charge at each entry, $x_{ij}$, is assumed to be made up by
an operator of higher dimensions with no hypercharge [\cite{FN}]. For
example, X-charge conservation allows the non-renormalizable term
$${\bf Q}_i\overline{\bf u}_jH_u\left({\theta\over M}\right)^{n_{ij}}\ ,
\eqno(nonren)$$
provided that  the $n_{ij}$ are positive numbers which satisfy
$$x_{ij}-xn_{ij}= 0 \ .\eqno(conserve)$$
We have introduced an electroweak singlet field $\theta$  with X-charge
$-x$, and $M$ is some large scale.  It is simplest  to  assume the
existence of only  one such field. In general the $n_{ij}$ are expected
to be integers, unless one is willing to envisage fractional powers of
the field $\theta$, stemming from non-perturbative effects.

In the simplest model, there is only one electroweak singlet chiral
superfield $\theta$, not chaperoned by its vectorlike partner.
Invariance under supersymmetry then naturally[\cite{LNS}]  generates a
true texture zero whenever a Yukawa matrix element has negative excess
$X$-charge in units of (-$x$), and non-zero entries correspond only to
positive excess X-charge. Henceforth we normalize X so that $x=1$.

In slightly more complicated models, $\theta$ is accompanied by its
vector-like partner $\overline\theta$, with opposite value of X-charge.
Any entry with negative excess charge can be filled by terms like
$${\bf Q}_i\overline{\bf u}_jH_u\left({\overline\theta\over M}
\right)^{n_{ij}}\ ,\eqno(nonren2)$$
showing that the excess charges  $x_{ij}$ need not be of the same sign.
This also  allows $\theta$ to have an X-conserving  mass.

Let us assume that  $\theta$ has a vacuum expectation value smaller
than M, producing a small parameter, $\theta/M$. The $n_{ij}$ then
determine the order of magnitude of the entries in the Yukawa
matrices[\cite{FN}]. The masses $M$, and thus the expansion parameters
are  in principle different in
the three charge sectors. However, since the down quark and lepton sectors
share the same electroweak quantum numbers, we expect them to be the
same at least for the charge -1 and -1/3 matrices.

This simple picture of the orders of magnitude of Yukawa matrices is quite
restrictive. Consider the general case, where the normalized
Yukawa matrix is
$$Y_{ij}={\cal O}(\lambda^{n_{ij}})\ ,\eqno(oom)$$
and
$$n_{ij}=\vert x_{ij}\vert\ ,\eqno(un)$$
normalized to the heaviest mass in each charge sector. From the
constraints satisfied among the $x_{ij}$
$$\eqalign{x_{11}&=x_{13}+x_{31}\ ,\qquad x_{22}=x_{23}+x_{32}\ ,\cr
x^{}_{12}&=x^{}_{13}+x^{}_{32}\ ,\qquad x^{}_{21}=x^{}_{23}+x^{}_{31}\
,\cr}\eqno(relations)$$
we obtain the inequalities
$$\eqalign{n^{}_{11}&\le n^{}_{13}+n^{}_{31}\ ;\qquad  n^{}_{22}\le n^{}_{23}
+n^{}_{32}\ ,\cr
n^{}_{12}&\le n^{}_{13}+n^{}_{32}\ ;\qquad  n^{}_{21}\le n^{}_{23}
+n^{}_{31}\ ,\cr }$$
These enable us to derive general results on the hierarchy of
eigenvalues of these matrices, in terms of
$$\eqalign{p&={\rm min}\ (n^{}_{11},n^{}_{22},n^{}_{12},n^{}_{21} )\ ,\cr
q&={\rm
min}\ (n^{}_{11}+n^{}_{22},n^{}_{12}+n^{}_{21}) \ .\cr}\eqno(pqequal)$$

By considering the characteristic equation of the hermitean combinations
${\bf Y}^\dagger{\bf Y} $ in each charge sector, we find the following
eigenvalue patterns
$$\eqalign{p\ge {q\over 2}\qquad
{\rm eigenvalues}&:~~{\cal O}(1)\ ,\ \pm{\cal O}(\lambda^p)
\ ,\cr p\le {q\over
2}\qquad{\rm eigenvalues}&:~~{\cal O}(1)\ ,\ {\cal O}(\lambda^p)\ ,
{\cal O}(\lambda^{q-p})\ .\cr}\eqno(patterns)$$
The first pattern is in  contradiction with data, leaving the second as
the only physically acceptable case. The determination of the order of
magnitudes has been reduced to that for the underlying $2\times 2$
Yukawa matrix. Let us note for further use that the geometric
hierarchies of the type \(top), \(bottom), \(lept) are obtained for
$q=3p$.

It is possible to classify the orders of magnitude of the eigenvalues,
according to the ranges taken by the X-charges in the different charge
sectors. When $a_3$ and $b_3$ have the same sign, the results can be
summarized as follows:

\noindent$\bullet$ $3\vert a_8+b_8\vert \ge \vert a_3+b_3\vert$
$$q=6\vert a_8+b_8\vert \ ;\qquad p=3\vert a_8+b_8\vert - \vert a_3+b_3\vert
\ ,\eqno(casesone)$$
with the geometric hierarchy for $\vert a_3+b_3\vert  =\vert
a_8+b_8\vert $.

\noindent$\bullet$ $
\vert a_3-b_3\vert \le 3 \vert a_8+b_8\vert \le \vert a_3+b_3\vert$
$$q=6\vert a_8+b_8\vert  \ ;~~p=\cases{3\vert a_8+b_8\vert - \vert a_3-b_3\vert
\ ,~{\rm when}~~6\vert a_8+b_8\vert\le \vert a_3-b_3\vert +\vert a_3+b_3\vert
\ ,\cr
\vert a_3+b_3\vert-3\vert a_8+b_8\vert
\ ,~{\rm when}~~6\vert a_8+b_8\vert\ge \vert a_3-b_3\vert +\vert
a_3+b_3\vert\ .\cr}\eqno(casestwo)$$
Geometric hierarchies for these two cases are possible only if $\vert
a_3+b_3\vert\ge 5\vert a_3-b_3\vert$, with the respective assignments $
\vert a_3-b_3\vert =\vert a_8+b_8\vert$ and $\vert a_3+b_3\vert =5\vert
a_8+b_8\vert$.

\noindent$\bullet$ $3\vert a_8+b_8\vert \le \vert a_3-b_3\vert$
$$q=2\vert a_3-b_3\vert \ ;\qquad p=\vert a_3-b_3\vert -3\vert a_8+b_8\vert
\ .\eqno(casesthree)$$
In this case the geometric hierarchy is obtained for $9\vert a_8+b_8\vert
=\vert a_3-b_3\vert$.

When $a_3$ and $b_3$ have opposite signs, we obtain the same equations
with $b_3$ replaced by $-b_3$. The other two charge sectors are
described by the same equations, by changing $b_i$ to $c_i$ for the down
quarks, and $a_i,b_i$ by $d_i,e_i$, respectively for the charged leptons
sector.

This analysis simplifies if restricted to  the case of symmetric
textures. A texture is said to be symmetric if $\vert x_{ij}\vert =\vert
x_{ji}\vert$, in a charge sector, which  does not necessarily mean that
the Yukawa matrices are symmetric, only their orders of magnitude. There
is no fundamental reason to require symmetry of the textures, although
it was found[\cite{RRR}] that several symmetric textures are compatible
with experiment. Symmetric textures that reproduce hierarchical
eigenvalues imply
$$\eqalign{{\rm Charge ~~{2\over 3}\  sector}:&\ \ \ a_3=b_3\ ,\qquad
a_8=b_8\ .\cr
{\rm Charge ~~-{1\over 3}\  sector}:&\ \ \ a_3=c_3\ ,\qquad
a_8= c_8\ .\cr
{\rm Charge ~~-1\  sector}:&\ \ \ d_3=e_3\ ,\qquad
d_8=e_8\ .\cr
}\eqno(symmetric)$$
The excess X-charge, shown here for a quark  Yukawa matrix, is
$$
\pmatrix{2\vert 3 a_8+a_3\vert  &6 \vert a_8\vert  & \vert 3a_8+a_3\vert \cr
6\vert  a_8\vert  &2\vert  3a_8-a_3\vert  & \vert 3a_8-a_3\vert  \cr
\vert  3a_8+a_3\vert & \vert 3a_8-a_3\vert  &0\cr}\ .$$
Our general analysis now reduces to just three cases. In all three,
$q=12\vert a_8\vert $, and
$$\eqalign{p=&2(3\vert a_8\vert -\vert a_3\vert)\ ,~{\rm when}~~3\vert
a_8\vert \ge
\vert a_3\vert\ ,\cr
p=&2(\vert a_3\vert -3\vert a_8\vert)\ ,~{\rm when}~~2\vert a_3\vert\ge
6\vert a_8\vert \ge \vert a_3\vert\ ,\cr
p=&6\vert a_8\vert\ ,~~{\rm when}~~\vert a_3\vert \ge 6\vert a_8\vert\ .\cr}
\eqno(symcases)$$
In the first case, geometric hierarchy is achieved for $\vert
a_3\vert=\vert a_8\vert\ ,$ corresponding to the U or V-spin of the
family $SU(3)$. The second case yields $\vert a_3\vert =5\vert a_8\vert
\ ,$ and the third case does not allow for a geometric hierarchy. In the
more constrained case of totally symmetric textures[\cite{RRR}],
equation \(goodrat) leads to
$$\vert a_8\vert =\vert b_8\vert =\vert c_8\vert =\vert d_8\vert =\vert
e_8\vert\ ,\eqno(allsym)$$
and geometric hierarchy in all sectors implies for all the textures
$$\vert a_3\vert ,\vert b_3\vert ,\vert c_3\vert,\vert d_3\vert,\vert
e_3\vert =\cases{~\vert a_8\vert\ ,\cr
5\vert a_8\vert\ .\cr}\eqno(allsym2)$$

{}From now on, we restrict our analysis to the case where
$\theta$ is chiral and all the excess charges have the same
sign. Then we always have $q=6(a_8+b_8)$, so that
$$\det{\bf
Y}_u={\cal O}(\lambda_u^{6(a_8+b_8)})\ .\eqno(det)$$
Hence the
X-charge of the determinant in each charge sector is {\it
independent} of the texture coefficients that distinguish
between the two lightest families. We set
$$\det {\bf Y}_u\sim
y_t^3 {\cal O}(\lambda_u^{U})\ ,\qquad \det {\bf Y}_d\sim y_b^3
{\cal O}(\lambda_d^{D})\ ,\qquad \det {\bf Y}_l\sim y_\tau^3
{\cal O}(\lambda_d^{E})\ ,\eqno(detall)$$
where
$$U\equiv
6(a_8+b_8)\ ,\ D\equiv 6(a_8+c_8)\ ,\  E\equiv 6(d_8+e_8)\ .$$
Since the down and lepton matrices have the same quantum
numbers, and couple to the same Higgs, we may assume they have
the same expansion parameter. In that case we can relate the
products of the down quark masses to that of the leptons
(assuming $y_b=y_\tau$)
$${ m_dm_sm_b\over m_em_\mu m_\tau}\sim
{\cal O}(\lambda_d^{(D-E)})\ . \eqno(detratio)$$
It is more
difficult to compare the up and down sectors in this way
because of the unknown  value of both $\tan\beta$, which sets
the normalization between the two sectors, and of  the relative
magnitudes of the expansion parameters. In terms of their
geometric mean and their ratio in the charge 2/3 and -1/3
sectors,
$$\lambda_0=\sqrt{\lambda_u\lambda_d}\ ,\qquad
\chi=\sqrt{{\lambda_u\over\lambda_d}}\ ,\eqno(params)
$$
we find
$${ m_um_cm_t\over m_dm_sm_b}\sim \tan^3\beta\left({y_t\over
y_b}\right)^3 {\cal O}(\lambda_0^{(U-D)}\chi^{(U+D)})\
.\eqno(detproduct)$$

Experimentally, this  ratio is known to be much larger than
one. If the expansion parameters are the same for all three
charge sectors, the data implies
$$U\approx
2D\approx 2E\approx 12\ ,\eqno(same)$$
The less model dependent  conclusion is that $D=E$, barring the
perverse possibility that numerically, $\lambda_d^{(E-D)}$
turns out to be of order one, which involves the overall
normalization of X. We cannot infer the value of $\tan\beta$
without assuming an order of magnitude for $\chi$ and for
$y_t/y_b$. From \(casesone), \(casestwo), \(casesthree), we see
that the hierarchy among the eigenvalues, however, does depend
on the charges of the first two families.

Finally, we note that the relations \(relations) imply testable order of
magnitude estimates among the Yukawa matrix elements. For example
$$Y_{11}\sim {Y_{13}Y_{31}\over Y_{33}}\ ,\qquad
Y_{22}\sim {Y_{23}Y_{32}\over Y_{33}}\ ,\eqno(yeses)$$
valid for each of the three charge sectors, and they are
consistent with many of the allowed textures[\cite{RRR}].
\vskip .5cm
\subhead{3.~\bf Anomalies}
\taghead{3.}
In general, the X family symmetry is anomalous. If it is not gauged,
this is not a cause for concern, although its spontaneous breakdown will
generate a massless familon[\cite{reiss}]. If gauged,
its anomalies must be accounted for. The three chiral families
contribute to the mixed gauge anomalies as follows
$$\eqalignno{C_3&=3(2a_0^{}+b_0^{}+c_0^{})\ ,&(anom3)\cr
C_2&=3(3a_0^{}+d_0^{})\ ,&(anom2)\cr
C_1&=a_0^{}+8b_0^{}+2c_0^{}+3d_0^{}+6e_0^{}\ .&(anom1)\cr}$$
The subscript denotes the gauge group of the Standard Model, {\it i.e.}
$1\sim U(1)$, $2\sim SU(2)$, and $3\sim SU(3)$.
The X-charge also has a mixed gravitational anomaly, which is simply
the trace of the X-charge,
$$C_g=3(6a_0^{}+3b_0^{}+3c_0^{}+2d_0^{}+e_0^{})+C_g^\prime\ ,\eqno(anomg)$$
where $C_g^\prime$ is the contribution from the particles that do not
appear in the minimal $N=1$ model.
One must also account for the mixed $YXX$ anomaly, given by
$$C_{YXX}=6(a_0^2-2b_0^2+c_0^2-d_0^2+e_0^2)+4A_T\ ,\eqno(anomixed)$$
with the texture-dependent part given by
$$A_T=(3a_8^2+a_3^2)-2(3b_8^2+b_3^2)+
(3c_8^2+c_3^2)-(3d_8^2+d_3^2)+(3e_8^2+e_3^2) \ .\eqno(AT)$$
The last anomaly coefficient is that of the X-charge itself, $C_X$,
the sum of the cubes of the X-charge.

Extra particles with chiral X-charge other than those in the minimal
model, will contribute to both $C^\prime_g$ and $C_X$, for instance,
right-handed neutrino partners of the charged leptons (left-right
symmetric theories). With only three chiral families (and not a fourth
with a massive neutrino), new particles with electroweak quantum numbers
must be electroweak vector-like pairs in order to have large $\Delta
I_w=0$ masses, but they need not be vector-like with respect to
X-charge, in which case they will contribute to the $C_i$ coefficients,
a possibility we do not address in this letter.

{}From the tree-level Yukawa couplings to the third family expressed
through \(Yuky), we can write
combinations of anomaly coefficients in terms of the family-dependent
charges
$$\eqalign{C_1+C_2-{8\over 3}C_3&=12(d_8+e_8-a_8-c_8)=2(E-D)\ ,\cr
C_3&=6(2a_8+b_8+c_8)=U+D\ .\cr}\eqno(anotext)$$
These  allow us to relate the anomaly
coefficients to the ratio of products of quark and lepton masses
\(detratio), (assuming $y_b=y_\tau$),
$$  {m_dm_sm_b \over m_em_\mu m_\tau}\sim {\cal
O}(\lambda_d^{-(C_1+C_2-8/3C_3)/2})\ .\eqno(detrat)$$
Compatibility with the extrapolated data requires the exponent to vanish
$$C_1+C_2-{8\over 3}C_3= 0\ ,\eqno(vanish)$$
which expressed in other variables, reads $E=D$. Another expression
relates the product of the six quark masses
$$\Pi\  m_q\sim  v^6(y_ty_b\sin\beta\cos\beta)^3{\cal
O}(\lambda_0^{C_3}\ \chi^{(n-p)})\ .\eqno(detpro)$$
If the expansion parameters are the same in both sectors ($\chi\approx
1$), then
$$C_3\approx 18\ .\eqno(c3s)$$
These equations apply only when all the $x_{ij}$ are all of the same
sign.

For the X-charge to be gauged, its anomalies must be cancelled. We
consider two  ways to achieve this. One is to arrange the
charges so that the anomalies cancel directly.  The second  is to
appeal to a Green-Schwarz mechanism. One could also add new
chiral fields to the minimal model to soak up the anomalies of the
minimal model fields, but we do not consider this complicated alternative.

Let us assume first that X is anomaly-free. Then we must have
$$C_1=C_2=C_3=0\ ,\qquad C^{}_g=0\ .\eqno(anofree) $$
The last equation is not constraining as there are likely more fields in
the theory with chiral X-charge. These are nicely consistent with
\(detrat), but the vanishing of $C_3$ contradicts our hypothesis that
all excess charges have the same sign. Indeed, using the tree-level
Yukawa relations \(Yuky), \(anom3), we see that
$$0=C_3=6(a_8+b_8)+6(a_8+c_8)\ ,$$
which is not consistent with our assumption that all excess charges are
positive. Hence we must rely on the Green-Schwarz mechanism.

\subhead{4.~\bf Green-Schwarz Cancellation of X Anomaly}
\taghead{4.}
If indeed, X is anomalous, we can appeal to the Green-Schwarz mechanism
to cancel some of its anomalies, and demand that the others vanish.
String theories naturally contain an antisymmetric tensor
Kalb-Ramond field. In four dimensions, it is the Nambu-Goldstone boson
of an anomalous $U(1)$ which couples like an axion through a dimension
five term to the divergence of the anomalous current. Its anomalies are
cancelled by the Green-Schwarz mechanism[\cite{GS}]. Under a chiral
transformation, this term is capable of soaking up certain anomalies, by
shifting the axion field, provided that they appear in commensurate
ratios
$${C_i\over k_i}={C_X\over k_X}={C_g\over k_g}\ ,\eqno(proportion)$$
where the $k_i$ are the Kac-Moody levels. They need to be integers only
for the non-Abelian factors. We have assumed that the mixed
gravitational anomaly is also cancelled {\it \`a la} Green-Schwarz. In
other theories, it could be cancelled in the traditional way.

In superstring theories, this $U(1)$ is broken spontaneously slightly
below the string scale. The scale is set by the charge content of the
theory[\cite{break}]. It follows that singlets with masses protected by
X can still be very massive, and not appear in the effective low-energy
theory.

This chiral U(1) charge may be useful for string phenomenology. Ib\` a\~
nez[\cite{Ib}] remarked that it can fix the value of the Weinberg
angle, without the use of a grand unified group. More recently, Ib\` a\~
nez and Ross[\cite{IR}] applied it to the determination of symmetric
textures. Following their approach, we investigate the constraints this
hypothesis puts on allowed textures at the string unification scale.

In superstring theories, the non-Abelian gauge groups have the same
Kac-Moody levels. For Green-Schwarz cancellation, it means that
$$C_2=C_3\qquad {\rm or}\qquad  d_0^{}=b_0^{}+c_0^{}-a_0^{}\ .\eqno(kac)$$
After this very generic requirement, we see that equation \(detrat)
reduces to
$$  {m_dm_sm_b \over m_em_\mu m_\tau}\sim {\cal
O}(\lambda_d^{-(C_1-5/3C_2)/2})\ ,\eqno(detrat2)$$
valid whenever $\theta$ is chiral. Since the right-hand side is of order
one, it means that  the exponent vanishes, so that in models
with an {\it ab initio} $\mu$ term, we {\it deduce} that
$$C_1={5\over 3}C_2\ .\eqno(good)$$
However the gauge coupling constants at string unification scale with
the anomaly coefficients, so that
$${C_1\over C_2}={g_1^2\over g_2^2}\ ,\eqno(weinberg)$$
which fixes the Weinberg angle to the value
$$\sin^2\theta_w={3\over 8}\ ,$$
at the string scale, in perfect agreement with extrapolated
phenomenology!  Hence with chiral $\theta$, and the $\mu$ term, the
Green-Schwarz cancellation leads to the correct value of the Weinberg
angle. This may be viewed as  a strong hint that the $N=1$ model does
indeed come from superstrings!
Alternatively we could have imposed the canonical Weinberg angle value
$$5C_2=3C_1\qquad {\rm or}\qquad e_0^{}=2a_0^{}-b_0^{}\ ,\eqno(angle)$$
which would have led us to agreement with experiment. From \(anotext),
we see that  $E=D$.

Equations \(kac) and \(angle), together with the tree level restrictions
\(Yuky),
allow us to express the family-independent charges, $X_0$, in terms of
the two observable combinations $U$, $E$, and the excess mixed
gravitational anomaly
$$\eqalign{
a_0^{}&={1\over 9}\big(5E+4U-C_g+C_g^\prime\big)\ ,\cr
b_0^{}&={1\over 9}\big(-5E-U+C_g-C_g^\prime\big)\ ,\cr
c_0^{}&={1\over 9}\big(-2E-4U+C_g-C_g^\prime\big)\ ,\cr
d_0^{}&=-{1\over 3}(4E+3U-C_g+C_g^\prime)\ ,\cr
e_0^{}&={1\over 3}(5E+3U-C_g+C_g^\prime)\ ,\cr
}\eqno(values)$$
We note that the gravitational anomaly is exactly along the anomaly-free
combination of baryon minus lepton numbers, $B-L$. In fact the most
general X-charge can contain an arbitrary mixture along $B-L$, but
this is already taken into account by our general parametrization.

The vanishing of the mixed anomaly  relates the excess mixed
gravitational anomaly to a combination of the family-dependent charges
$$C^{}_g-C^\prime_g={5\over 2}(U+2E)+{9A_T\over 2(U+E)}\
,\eqno(grave)$$
where $A_T$ is the previously defined family-dependent part.
We can use this equation to express the family-independent charges in
terms of $U$, $E$, and $A_T$, with the simple results
$$\eqalign{a_0^{}={U\over 6}-{A_T\over 2(U+E)}\ ,&\qquad
b_0^{}={U\over 6}+{A_T\over 2(U+E)}\ ,\cr
c_0^{}={E\over 3}-{U\over 6}+{A_T\over 2(U+E)}\ ,&\qquad
d_0^{}={E\over 3}-{U\over 6}+{3A_T\over 2(U+E)}\ ,\cr
e_0^{}={U\over 6}-&{3A_T\over 2(U+E)}\ ,\cr}\eqno(assign)$$
In superstring models, the Green-Schwarz mechanism extends to the mixed
gravitational anomaly so that
$${C_g\over C_3}={k_g\over k_3}=\eta\ .$$
where $\eta$ is a normalization parameter; in the simplest
level-one models, it is equal to $12$. In general, however,
$$C_g=\eta(U+E)\ .\eqno(gsmixed)$$
The family independent X-charges are seen to depend only on two
parameters, $E$, and $U$, assuming we know the normalization $\eta$.

\subhead{5.~\bf Results}
\taghead{5.}

We have seen that when all the excess X-charges are of the same sign,
the family symmetry must be anomalous in order to produce textures in
agreement with experiment. When the anomalies are compensated by the
Green-Schwarz mechanism, coming from superstring theory, we find that
the near equality of the products of charged lepton and down quark
masses in the ultraviolet \(detrat) fixes the Weinberg angle to be
$\sin^2\theta_w={3\over 8}\ ,$ in perfect agreement with experiment.

Consider first a simple model where  the textures are symmetric in all
three charge sectors. This means that
$$a_{3,8}=b_{3,8}=c_{3,8}\ ,\qquad d_{3,8}=e_{3,8}\ ,$$
so that  $U=E$, and  $A_T=0$. The mixed gravitational anomaly
is fixed to be
$$C^{}_g-C^\prime_g={15\over 2}U\ , $$
and  from \(assign), the family-independent charges reduce to
$$a_0^{}=b_0^{}=c_0^{}=d_0^{}=e_0^{}={U\over 6}\ .$$
Geometric hierarchy in all three sectors can be achieved by
choosing X to be  the third component of the U-spin
subgroup of the family $SU(3)$
$$a_8^{}=a_3^{}=b_3^{}=c_3^{}=d_3^{}\ .$$
However the expansion parameter
cannot be the same in the charge 2/3 and -1/3 sectors. Since
$U=D$, comparison of \(top), \(bottom), and \(detall) implies
that  $\lambda_u^{}=\lambda_d^2$, which means that
$\lambda_d=M_d/M_u$, where $M_{u,d}$ are the mass
parameters  in the non renormalizable interactions \(nonren)and
\(nonren2). If the expansion parameters are all the same, the three
Yukawas cannot have symmetric textures.

If we want to have the same expansion parameter in all three sectors, we
have to relax this constraint by supposing that the textures are
symmetric only for the down quark and charged lepton sectors. Then we
can have the same expansion parameter in the two quark sectors, by
requiring that $U=2E$, that is $b_8=3a_8$. A particularly simple
assignment is to take
$$a_8^{}=c_8^{}=d_8^{}=e_8^{}=a_3^{}=c_3^{}=d_3^{}=e_3^{}={b_8\over
3}={b_3\over 3}\ .$$
This leaves us with a geometric hierarchy in all three sectors,
and the same expansion parameter. The order of eigenvalues are
$${\cal O}(1)\ ,\ {\cal O}(\lambda^{8a_8})\ ,\ {\cal
O}(\lambda^{16a_8})\ ,$$
for the charge 2/3 sector, and
$${\cal O}(1)\ ,\ {\cal O}(\lambda^{4a_8})\ ,\ {\cal
O}(\lambda^{8a_8})\ ,$$
for the charge -1/3 and -1 sectors. The family independent X-charges are
different in all charge sectors
$$X_0=({2\over 3})^2a_8(11,7,-2,-6,15)\ ,$$
in the notation of \(Xfamily). However, if we subtract an appropriate
amount of hypercharge, we find a much simpler assignment,
$$X^\prime_0=({2\over 3})^2{3a_8\over 5}(17,17,-6,-6,16)\ ,$$
yielding the same X-charge for members of the same $SU(5)$ multiplets.
The family-dependent part is along the third component of family U-spin,
to reproduce the geometrical hierarchies.

There are clearly many other possible charge assignments which
reproduce the geometric hierarchies and yield the same
expansion parameter in all three sectors. To choose
among them requires  making detailed assumptions about the origin of
the non-renormalizable operators, and the interaction of the
electroweak singlet. It was not the purpose of this letter to
offer a detailed model, but to show that many of the generic
features in the data can be understood simply in terms of a
gauged family Abelian symmetry. In the process, we have been
able to relate observable quantities to its anomalies, with the
startling result that the value of the Weinberg angle is fixed
by simply demanding that the ratio of down quark to lepton
masses be of order one!

We leave to a future publication[\cite{BLR}] the analysis of more
general models, where the electroweak singlet field is accompanied by
its vector-like partner, and where the $\mu$ term restriction
is lifted.

One of us (P.R.) wishes to thank Professors T. Banks, J. Harvey
and Y. Nir for useful comments, and the Aspen Center for
Physics for its hospitality, where part of this work was done.
We also wish to thank St\'ephane Lavignac for helpful
discussions.

\references

\refis{unification} P.~Langacker, in Proceedings of the
PASCOS90 Symposium, Eds.~P.~Nath
and S.~Reucroft, (World Scientific, Singapore 1990).

\refis{FN} C.~Froggatt and H.~B.~Nielsen \np B147, 277, 1979.

\refis{reviews}
For reviews, see H.~P.~Nilles,  \prpts 110, 1, 1984 and
H.~E.~Haber and G.~L.~Kane, \prpts 117, 75, 1985.

\refis{gut}
J.~C.~Pati and A.~Salam,
\pr D10, 275, 1974;
H.~Georgi and S.~Glashow,
\prl 32, 438, 1974;
H.~Georgi, in {\it Particles and Fields-1974}, edited by C.E.Carlson,
AIP Conference Proceedings No.~23 (American Institute of Physics,
New York, 1975) p.575;
H.~Fritzsch and P.~Minkowski,
\journal Ann.~Phys.~NY, 93, 193, 1975;
F.~G\" ursey, P.~Ramond, and P.~Sikivie,
\pl 60B, 177, 1975.

\refis{btau} H.~Arason, D.~J.~Casta\~no, B.~Keszthelyi, S.~Mikaelian,
E.~J.~Piard, P.~Ramond, and B.~D.~Wright,
\prl 67, 2933, 1991;
A.~Giveon, L.~J.~Hall, and U.~Sarid,
\pl 271B, 138, 1991.

\refis{RRR}P. Ramond, R.G. Roberts and G. G. Ross, \np B406, 19, 1993.

%\refis{Kusenko} A. Kusenko and R. Shrock, \pr D50, 30, 1994.

\refis{Ib}L. Ib\'a\~nez, \pl B303, 55, 1993.

\refis{IR}L. Ib\'a\~nez and G. G. Ross, \pl B332, 100, 1994.

\refis{unification}
U.~Amaldi, W.~de Boer, and H.~Furstenau,
\pl B260, 447, 1991;
J.~Ellis, S.~Kelley and D.~Nanopoulos,
\pl 260B, 131, 1991;
P.~Langacker and M.~Luo,
\pr D44, 817, 1991.

\refis{reiss} D. B. Reiss, \pl B155, 217, 1982;
F. Wilczek, \prl 49, 1549, 1982.

\refis{hall} S. Dimopoulos, L. Hall, S. Raby, and G. Starkman, \pr
D49, 3660, 1994.

\refis{LNS} M. Leurer, Y. Nir, and N. Seiberg, \np B398, 319, 1993.

\refis{GS} M. Dine, N. Seiberg, and E. Witten, \np B289, 585, 1987;
J. Atick, L. Dixon, and A. Sen, \np B292, 109, 1087;
M.  Dine, I. Ichinoise, and N. Seiberg, \np B293, 253, 1987.

\refis{break} A. Font, L.E. Ib\' a\~ nez, H. P. Nilles, and F. Quevedo,
\np B307, 109, 1988; \pl B210, 101, 1988;
J. A. Casas, E. K. Katehou, and C. Mu\~ noz, \np B317, 171, 1989;
J. A. Casas, and C. Mu\~ noz, \pl B209, 214, 1988; \pl B214, 63, 1988;
A. Font, L.E. Ib\' a\~ nez, F. Quevedo, and A. Sierra, \np B331, 421,
1990.

%\refis{muterm} J.E. Kim and H. P. Nilles, \pl B138, 150, 1984.

%\refis{nilles} J.E. Kim and H. P. Nilles, \pl B263, 79, 1991; E. J.
%Chun, J.E. Kim and H. P. Nilles, \np B370, 105, 1992.

%\refis{camu} J. Casas and C. Mu\~noz, \pl B306, 288, 1993.

%\refis{giuma} G. Giudice and A. Masiero, \pl B206, 480, 1988.

\refis{seesaw}M. Gell-Mann, P. Ramond, and R. Slansky in Sanibel
Talk,
CALT-68-709, Feb 1979, and in {\it Supergravity} (North Holland,
Amsterdam 1979). T. Yanagida, in {\it Proceedings of the Workshop
on Unified Theory and Baryon Number of the Universe}, KEK, Japan,
1979.

\refis{wolf} L. Wolfenstein, \prl  51, 1945, 1983.

\refis{BLR} P. Bin\'etruy, S. Lavignac, and P. Ramond, in preparation.

\refis{PAPAG} E. Papageorgiou, $``$Yukawa Textures from an extra $U(1)$
Symmetry", Orsay Preprint, LPTHE Orsay 40/94.

\refis{WORKSHOP} We reported our preliminary results at the Fermilab
workshop on Yukawa Couplings, Nov., 1994, where we learned that a
similar line of enquiry was being followed by V. Jain and R. Shrock.

\endreferences\endit
\end